\documentclass[12pt]{article}

\usepackage{amsmath,amssymb,amsthm,amsxtra,overpic,bbm,bm,epsfig,subfigure}
\usepackage{cite}
\usepackage{graphicx}
\usepackage{hyperref}
\usepackage[margin=2.5cm]{geometry}
\usepackage{stackengine}
\usepackage{pifont} 
\usepackage{booktabs}
\usepackage{pstricks}
\usepackage{color}
\usepackage{slashed}
\usepackage[capitalise]{cleveref}
\usepackage{enumerate}
\usepackage{bbold}

\everymath{\displaystyle}

\newcommand{\FL}{\Delta_{F_L}}
\newcommand{\DF}{\Delta_F}
\newcommand{\FR}{\Delta_{F_R}}
\newcommand{\DC}{\Delta_\chi}

\DeclareMathOperator{\Tr}{Tr}

\begin{document}
\vspace{6mm}

\begin{center}
{\Large\bf Axionic Dirac seesaw and\\[2mm] electroweak vacuum stability}\\
\vspace{0.2cm}
\end{center}

\vspace{0.1cm}

\begin{center}
{\bf J. T. Penedo}$^{\,a,}$\footnote{Email: \texttt{joao.t.n.penedo@tecnico.ulisboa.pt}},
\quad
{\bf Yakefu Reyimuaji}$^{\,b,}$\footnote{Email: \texttt{yreyi@hotmail.com}},
\quad
{\bf Xinyi Zhang}$^{\,c,}$\footnote{Email: \texttt{zhangxinyi@ihep.ac.cn}}
\\
\vspace{0.2cm}
$^a$ {\em \small Centro de Física Teórica de Partículas, CFTP, Departamento de Física,\\
Instituto Superior Técnico, Universidade de Lisboa,\\
Avenida Rovisco Pais nr.~1, 1049-001 Lisboa, Portugal;}\\
$^b$ {\em \small School of Physical Science and Technology,
Xinjiang University, Urumqi, Xinjiang 830046, China}\\
$^c$ {\em \small Institute of High Energy Physics, Chinese Academy of Sciences, Beijing 100049, China}\\
\end{center}
\setcounter{footnote}{0}

\begin{abstract}
We explore the connection between tree-level Dirac neutrino masses and axion physics in a scenario where the PQ symmetry enforces lepton number conservation perturbatively.
Requiring that the PQ scale $f_a$ is the only heavy scale to play a role in neutrino mass generation, we are led to the construction of a KSVZ-type model where Dirac neutrino masses are inversely proportional to $f_a$, provided a real scalar triplet (zero hypercharge) is added to the SM scalar sector.
We analyse this extended scalar sector, focusing on the stabilisation of the electroweak vacuum.
The contribution of the triplet VEV to the $W$ mass may also be responsible for the recent hint of beyond-the-SM physics by the CDF collaboration.
\end{abstract}

\section{Introduction}

Despite its many successes, the Standard Model (SM) cannot be a final description of Nature. It must be extended in order to clarify the origins of neutrino masses and dark matter. Moreover, the SM by itself does not offer an explanation to the non-observation of the neutron electric dipole moment. This so-called strong CP problem can be solved via the Peccei-Quinn (PQ) mechanism~\cite{Peccei:1977hh,Peccei:1977ur,Wilczek:1977pj,Weinberg:1977ma}, whereby an axion is introduced in the theory (for a recent review see~\cite{DiLuzio:2020wdo}). From the point of view of a UV completion, this QCD axion arises as the pseudo-Nambu-Goldstone boson of a spontaneously broken, anomalous $U(1)_\text{PQ}$ symmetry.

Depending on its properties, the axion can provide the desired dark matter candidate. On the other hand, axion physics may directly connect to the generation of neutrino masses. In the case of Majorana neutrinos, the PQ scale $f_a$ is naturally identified with the type-I seesaw scale~\cite{Kim:1981jw,Langacker:1986rj,Shin:1987xc,Dias:2014osa,Ballesteros:2016euj}. The lepton-number-violating right-handed (RH) neutrino Majorana mass term is thus generated from a coupling of the type $\sigma \,\overline{\nu_R} \,\nu_R^c$, where $\sigma$ is the PQ scalar field. In this case, light neutrino masses are suppressed by the PQ scale, $m_\nu \sim v^2 / f_a$, with $v \simeq 246$ GeV.

At present, the nature of neutrino masses is not known. Dirac neutrinos remain a viable and interesting possibility. In this context, however, the connection to axion physics is not so direct as in the Majorana case. This link has been explored in models where Dirac neutrino masses are generated at the tree level~\cite{Gu:2016hxh, Baek:2019wdn, Peinado:2019mrn,delaVega:2020jcp,Dias:2020kbj,Berbig:2022pye} and at the one-loop level~\cite{Chen:2012baa, Carvajal:2018ohk,delaVega:2020jcp,Carvajal:2021fxu}.%
\footnote{
The baryon asymmetry of the universe may be generated in such a setup by the neutrinogenesis mechanism (aka Dirac leptogenesis)~\cite{Dick:1999je,Murayama:2002je}, see e.g.~\cite{Gu:2016hxh}.}
Focusing on the tree-level case, one typically finds $m_\nu \sim v f_a / \Lambda_\text{UV}$~\cite{Peinado:2019mrn}, i.e.~neutrino masses are proportional to the PQ scale and inversely proportional to an (in general) unrelated scale of new physics $\Lambda_\text{UV}$, e.g.~the GUT scale. Here, the suppression of $m_\nu$ with respect to the electroweak scale arises from the smallness of Yukawa couplings and of the ratio $f_a/\Lambda_\text{UV}$.
If an additional mass scale $\mu$ is present in the theory, one can instead obtain a relation of the type $m_\nu \sim \mu v f_a / \Lambda_\text{UV}^2$~\cite{Gu:2016hxh,Baek:2019wdn,delaVega:2020jcp}, with the ratio $\mu/\Lambda_\text{UV}$ possibly providing a further source of suppression.

\begin{figure}[t]
\centering
\includegraphics[width=\textwidth]{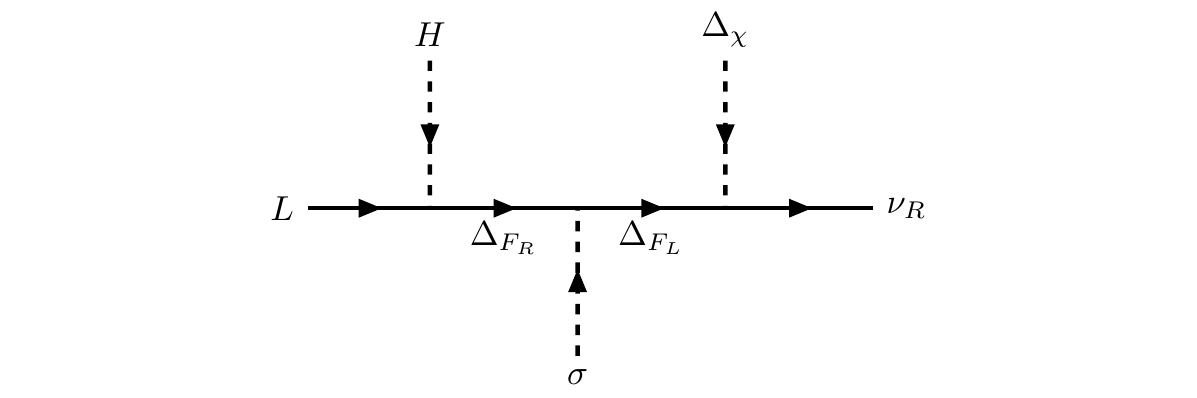}
\caption{Tree-level Feynman diagram giving rise to PQ-suppressed Dirac neutrino masses. The seesaw partners $\FL$ and $\FR$ have masses proportional to $f_a$, enabling the identification of the (Dirac) seesaw and PQ scales.
}
\label{fig:feynman}
\end{figure}

In this work, we look into the possibility of identifying the Dirac seesaw scale with the PQ scale $f_a$, so that the Dirac neutrino masses are suppressed by $f_a$ as in the Majorana case.%
\footnote{Recently Ref.~\cite{Berbig:2022pye} appeared, where the relation $m_\nu \sim v^3/f_a^2$ for Dirac neutrino masses is obtained.}
To avoid introducing an independent heavy scale $\Lambda_\text{UV}$, we focus on the diagram of \Cref{fig:feynman} as the main contribution to neutrino masses, which effectively corresponds to a dimension-5 operator.
In this case, one obtains a relation of the type $m_\nu \sim \mu v / f_a$, where $\mu$ corresponds to the vacuum expectation value (VEV) of a new neutral scalar. Such a relation was also found in the 3-3-1 setup of Ref.~\cite{Dias:2020kbj}, with $\mu = 10^4$ GeV.

Moreover, we find that the PQ symmetry by itself is enough to explain the Dirac nature of neutrino masses in such a setup. Namely, one does not need to impose an additional lepton-number symmetry, since PQ charges forbid Majorana mass terms at all perturbative orders. Such an economical possibility was previously explored in Refs.~\cite{Peinado:2019mrn,delaVega:2020jcp} for different classes of models. Unlike these models, which consider Dine-Fischler-Srednicki-Zhitnitsky (DSFZ)-type axions~\cite{Dine:1981rt,Zhitnitsky:1980tq}, we develop a scenario where the scalar fields (apart from $\sigma$) are not charged under $U(1)_\text{PQ}$. Thereby the SM Higgs doublet is not charged under this symmetry (neither are the SM quarks) and our axion is of the Kim–Shifman–Vainshtein–Zakharov (KSVZ) type~\cite{Kim:1979if,Shifman:1979if}. The considered extension of the scalar sector naturally warrants an analysis of the stability of the electroweak vacuum.

In \cref{sec:framework} we describe our framework, detailing the field content and neutrino masses. We further comment on the solution to the strong CP problem and on the possibility of explaining the recent CDF anomaly due to the contribution of the new scalar VEV to the $W$ boson mass. In \cref{sec:stability} we analyse the scalar sector of the theory. In particular, we look into the constraints imposed by vacuum stability on the discussed model. Finally, we present our conclusions in \cref{sec:summary}.

\section{Framework}
\label{sec:framework}

\subsection{Axionic Dirac seesaw}
\label{sec:numasses}
\subsubsection{Field content}
\label{sec:fcontent}
We start by setting the field content. Aiming at identifying the seesaw scale with the PQ-breaking scale, i.e.~$\Lambda_\mathrm{seesaw} \sim f_a$, first of all we introduce one complex singlet PQ field $\sigma$.
Maintaining a minimal field content, we introduce 2 generations of RH neutrinos $\nu_R$ and are led to the dimension-5 operator of the form $\overline{\nu_R} L H \chi /f_a$, where $L$ and $H$ are the SM lepton and Higgs doublets, respectively. Here, $\chi$ can be either a singlet scalar or an $SU(2)_L$ triplet, with zero hypercharge.
In order to generate the Dirac neutrino masses at tree level, we open up the dimension-$5$ operator by introducing a vector-like fermion, $\DF=\FR+\FL$ which gains a mass after PQ-symmetry breaking, i.e.~$m_{\DF} \sim f_a$. This corresponds to the seesaw diagram shown in \Cref{fig:feynman}.

Going forward, we consider the case when $\chi = \DC$ is an $SU(2)_L$ triplet, whereas the singlet possibility will be explored elsewhere~\cite{PRZ}.%
\footnote{Note that to preserve the structure of the diagram in \Cref{fig:feynman} in the absence of extra symmetries, the singlet would have to carry a non-zero PQ charge and thus be complex, in order to forbid a direct Majorana-type $\overline{\nu_R^c} \FR \sigma $ coupling.}
The minimal choice then corresponds a real $Y=0$ scalar triplet (see e.g.~\cite{Blank:1997qa,Forshaw:2001xq,Forshaw:2003kh,Khan:2016sxm}) instead of a complex one.
It follows that the fermion fields $\FR$ and $\FL$ must also be triplets. We are thus dealing with a type-III Dirac seesaw, in the terminology of Refs.~\cite{Gu:2016hxh,CentellesChulia:2018gwr}.
The triplets of the model are defined as
\begin{align}
	\DC\equiv \frac{1}{2}\begin{pmatrix}
		\chi^0 & \sqrt{2} \chi^+ \\
		\sqrt{2} \chi^- & -\chi^0
	\end{pmatrix},\,\, 
	\FR\equiv \frac{1}{2}\begin{pmatrix}
	 F^0_R  & \sqrt{2}  F^+_R \\
	 \sqrt{2} F^-_R & -F^0_R
   \end{pmatrix},\,\,
   \FL\equiv\frac{1}{2}\begin{pmatrix}
        F^0_L &  \sqrt{2} F^+_L \\
        \sqrt{2} F^-_L & - F^0_L
    \end{pmatrix}\;,
\label{eq:triplets}
\end{align}
where $\chi^0$ is real and $(\chi^+)^*=\chi^-$. Notice that $\DC=\DC^\dagger$. For $\FR$ and $\FL$, the component fields are all complex and $(F^+_{R,L})^* \neq F^-_{R,L}$. 

The leading-order contribution to the light neutrino mass scale can be read from the considered diagram. One has
\begin{align}
    m_\nu 
    \sim \frac{v_\chi\, v}{v_\sigma} \;,
\label{eq:dss}
\end{align}
where we assume that the neutral components of the scalars all acquire VEVs, i.e.~$\langle H^0 \rangle = v/\sqrt{2}$, $\langle \chi^0 \rangle =v_\chi$, and $\langle \sigma \rangle=v_\sigma/\sqrt{2} = \sqrt{2} N f_a$, with $N$ being the QCD anomaly coefficient. 
Note that at least 2 copies of the vector-like fermions $\DF$ are required to generate both $\Delta m^2_\odot$ and $\Delta m^2_\text{atm}$ neutrino mass-squared differences.
To obtain a sub-eV mass for the light neutrinos, we need $v_\chi/v_a \sim 10^{-12}$.
It is curious that the experimental constraints on $v_\chi$, suggesting $v_\chi \sim \mathcal{O}(\text{GeV})$ at most  (see~\cref{sec:W}), together with a typical scale of $f_a \sim 10^{9}\,\text{--}\,10^{12}$ GeV for axionic dark matter leads to viable neutrino mass scales for $\mathcal{O}(10^{-3}\,\text{--}\,1)$ Yukawa couplings.

Finally, to address the strong CP problem we introduce a vector-like quark $Q=Q_L+Q_R$, such that the SM quarks need not be charged under the PQ symmetry. Our model is therefore of the KSVZ-type (see also~\cref{sec:axion}).

\subsubsection{PQ as a lepton number symmetry}
Having set the field content, we show that it is possible to impose no other symmetry aside from the PQ symmetry --- especially no independent  global lepton number symmetry --- to guarantee Diracness. This requires that we charge the fields properly. To start, the PQ field is charged PQ$(\sigma)=1$, while PQ$(Q_{L,R}) = \pm 1/2$ for the vector-like quark, as usual. Since we work in a KSVZ-type model, the SM Higgs is not charged under the PQ symmetry and, consequently, the SM lepton doublet needs to carry a charge PQ$(L)\equiv\alpha \neq 0$ to forbid a Weinberg-operator contribution to light neutrino masses. 
The charge assignments for all relevant fields are collected in \Cref{tab:charges_triplet}. They follow from requiring that the interactions contained in the diagram of \Cref{fig:feynman} are allowed. One has PQ$(\Delta_\chi) = 0$, since it is a real scalar triplet.

\begin{table}[t]
  \centering
  \begin{tabular}{lcccccccccccc}
    \toprule
   Field & &   $L$ & $e_R$ & $\nu_R$ & $\FR$ & $\FL$& $Q_R$ & $Q_L$ & & $H$ & $\sigma$ & $\DC$\\
    \midrule
$SU(3)_c$ &&  $\mathbf{1}$  & $\mathbf{1}$     &  $\mathbf{1}$      &     $\mathbf{1}$          &     $\mathbf{1}$         &    $\mathbf{3}$  &   $\mathbf{3}$   &&  $\mathbf{1}$  &  $\mathbf{1}$       &  $\mathbf{1}$  \\[2mm]
$SU(2)_L$ &&  $\mathbf{2}$  & $\mathbf{1}$     &  $\mathbf{1}$      &     $\mathbf{3}$          &     $\mathbf{3}$         &    $\mathbf{1}$  &   $\mathbf{1}$   &&  $\mathbf{2}$  &  $\mathbf{1}$       &  $\mathbf{3}$  \\[2mm]
$U(1)_Y$  && $-\frac{1}{2}$ & $-1$ & 0  &  0    & 0  &     0 &  0          &&$\frac{1}{2}$ & 0       & 0   \\[2mm]
$U(1)_\text{PQ}$ && $\alpha$ & $\alpha$ & $\alpha+1$& $\alpha$ & $\alpha+1$& $-\frac{1}{2}$ &   $\frac{1}{2}$  && 0 & 1 & 0       
\\
    \bottomrule
  \end{tabular}
  \caption{Charge assignments of the considered Dirac seesaw model ($\alpha \neq 0$).}
  \label{tab:charges_triplet}
\end{table}

Note that the direct Dirac coupling $\bar{L} \tilde{H} \nu_R$ is automatically forbidden. To ensure that light neutrino masses are generated by the described Dirac seesaw mechanism, one also needs to forbid other possible Majorana contributions. This puts some additional constraints on the PQ charge $\alpha$, namely
\begin{enumerate}[i)]
    \item $\alpha \neq -1$ to avoid a direct RH neutrino Majorana mass term $\overline{\nu_R^c} \nu_R$, automatically forbidding higher-dimensional $\overline{\nu_R^c} \nu_R \DC^n$ terms ($n$ even, to form $SU(2)_L$ singlets), and more generally
    \item $\alpha \neq k/2$ ($k\in \mathbb{Z}$) to forbid (possibly higher-dimensional) $\overline{\nu_R^c} \nu_R (\sigma^{(*)})^n$ Majorana terms and their variants with additional $H$ or $\DC$ insertions.
\end{enumerate}
This last requirement of $2\alpha \notin \mathbb{Z}$ contains the previous ones. It forbids Weinberg-operator contributions with any number of $\sigma^{(*)}$ insertions.
Other possible Majorana-like contributions such as $\overline{\FR^c}\FR$, $\overline{\FL^c}\FL$, and $\overline{\FR^c}\nu_R\DC^n$ ($n$ odd) are also not allowed, even with an arbitrary number of $\sigma^{(*)}$ insertions, since these carry integer PQ charge. 

Making, for definiteness, the choice $\alpha= -1/3$ (in a parallel with SM quark e.m.~charges), one finds that neutrinos are Dirac particles in this model. Lepton number conservation is hence enforced (perturbatively) by the PQ symmetry.

The relevant Lagrangian $\mathcal{L}$, extending the SM one, is
\begin{subequations}
\begin{align}
    &\mathcal{L} =\mathcal{L}_\mathrm{kin} - \mathcal{L}_\mathrm{Yuk} -V(H,\DC,\sigma)\;,\\
    &\mathcal{L}_\mathrm{kin} =\left|\partial_\mu \sigma\right|^2 
    + \mathrm{Tr} \left| D_\mu \DC \right|^2
    + \overline{\nu_R} i \slashed{\partial} \nu_R
    + \overline{Q} i \slashed{D} Q
    + \overline{\DF} i \slashed{D} \DF
    \;,\\
    &\mathcal{L}_\mathrm{Yuk} = Y_Q\, \overline{Q_L} Q_R\sigma+ \overline{L} \tilde{H} \,Y_L\, \FR  +    \Tr \left(\overline{\FL} \,Y_F\,\FR \right) \sigma
    +  \Tr \left(\overline{\FL} \DC^* \right) \,Y_R \,\nu_R + \mathrm{H.c.}\;,
\end{align}
\end{subequations}
where the covariant derivative for the zero-hypercharge triplets $\Delta$ of \cref{eq:triplets} acts as $D_\mu \Delta=\partial_\mu \Delta + i g_2 [W_\mu, \Delta]$, with $W_\mu = W_\mu^a T_a$ containing the $SU(2)_L$ gauge bosons $W_\mu^a$ and $g_2$ being the corresponding gauge coupling.
Here, $Y_Q$ is a Yukawa coupling, while $Y_L$, $Y_F$, and $Y_R$ are Yukawa coupling matrices. In the minimal setup, $Y_L$ is a $3\times 2$ matrix, while $Y_F$ and $Y_R$ are $2\times 2$ matrices.

The scalar potential reads
\begin{align}
V(H,\sigma, \DC) = & -\mu_H^2 H^\dagger H -\mu^2_\chi \Tr \left(\DC^2\right)-\mu^2_\sigma \sigma^* \sigma +\kappa H^\dagger \DC H \nonumber \\
&
+ \frac{\lambda}{2} (H^\dagger H)^2 + \frac{\lambda_{\chi}}{2} \Tr\left(\DC^4\right)  + \frac{\lambda_\sigma}{2} (\sigma^* \sigma)^2\nonumber \\
					& +\frac{\lambda_{a}}{2}  (H^\dagger H)\Tr\left(\DC^2\right)
					+\frac{\lambda_{b}}{2} (H^\dagger H)(\sigma^* \sigma)   +\frac{\lambda_{c}}{2}  (\sigma^* \sigma)\Tr\left(\DC^2\right) \;,
\label{eq:potential-sim}
\end{align}
where all the couplings are real. Note also that $\DC^2 = \left[\left((\chi^0)^2/2+\chi^+\chi^-\right)/2\right] \mathbb{1}$, directly implying that the terms $\left[\Tr\left(\DC^\dagger \DC\right)\right]^2 = 2 \Tr\left(\DC^4\right)$ and $H^\dagger \DC^\dagger \DC H = (H^\dagger H)\Tr\left(\DC^2\right)/2$ are not new.
As is mentioned in Ref.~\cite{FileviezPerez:2008bj}, in the limit of vanishing $\kappa$ the potential possesses a global symmetry $O(4)_H\times O(3)_{\DC}$ and a discrete symmetry $\DC \rightarrow - \DC$.%
\footnote{Our potential also has an additional $O(2)_\sigma$ global symmetry.}
Thus $\kappa$ is protected by these symmetries and a small but non-vanishing $\kappa$ corresponds to their soft breaking. One may fix the sign of $\kappa$ via a sign flip of $\Delta_\chi$ and $\nu_R$. We consider the convention $\kappa > 0$ in what follows.

\subsection{The axion and the solution to the strong CP problem}
\label{sec:axion}
The solution to the strong CP problem in our model is identical to that of the KSVZ model~\cite{Kim:1979if,Shifman:1979if}, which we briefly review here. Anticipating the spontaneous breakdown of the PQ symmetry, one can parameterise the PQ field as
\begin{align}
    \sigma=\frac{1}{\sqrt{2}} \left( v_\sigma + \rho_\sigma \right) e^{i a/v_\sigma} \;,
\end{align}
where $a$ is the Goldstone field, i.e.~the axion, and $\rho_\sigma$ is the radial mode. The vector-like quark gets a mass from its interaction with the PQ field, $m_Q= Y_Q v_\sigma e^{i a/v_\sigma}/\sqrt{2}$. By performing an axial transformation $Q\rightarrow e^{-i \gamma_5 a/(2 v_\sigma)} Q$, the field-dependent phase in $m_Q$ gets rotated away. The $Q$ field is thus disentangled from the axion field and can be integrated out. The axial transformation is anomalous, leading to the $a G \tilde{G} /v_\sigma$ term, where $G$ is the $SU(3)_c$ field strength tensor. This term can be used to cancel the $\theta$ term, in a dynamical solution to the strong CP problem.

Comparing the generated $a G \tilde{G}$ term to the corresponding one in the axion effective Lagrangian, it follows that $v_\sigma = 2 N f_a$, as indicated in \cref{sec:fcontent}. Note that there are fields other than $\sigma$ that are charged under the PQ symmetry.
The QCD anomaly coefficient is $N=1/2$ in our model, and thus $v_\sigma = f_a$. It is more transparent to look back at \cref{eq:dss} with $v_\sigma = f_a$.
Since the domain wall number is $N_{DW}=2N=1$, this model is free from the cosmological domain wall problem.

\begin{figure}[t]
\centering
\includegraphics[width=\textwidth]{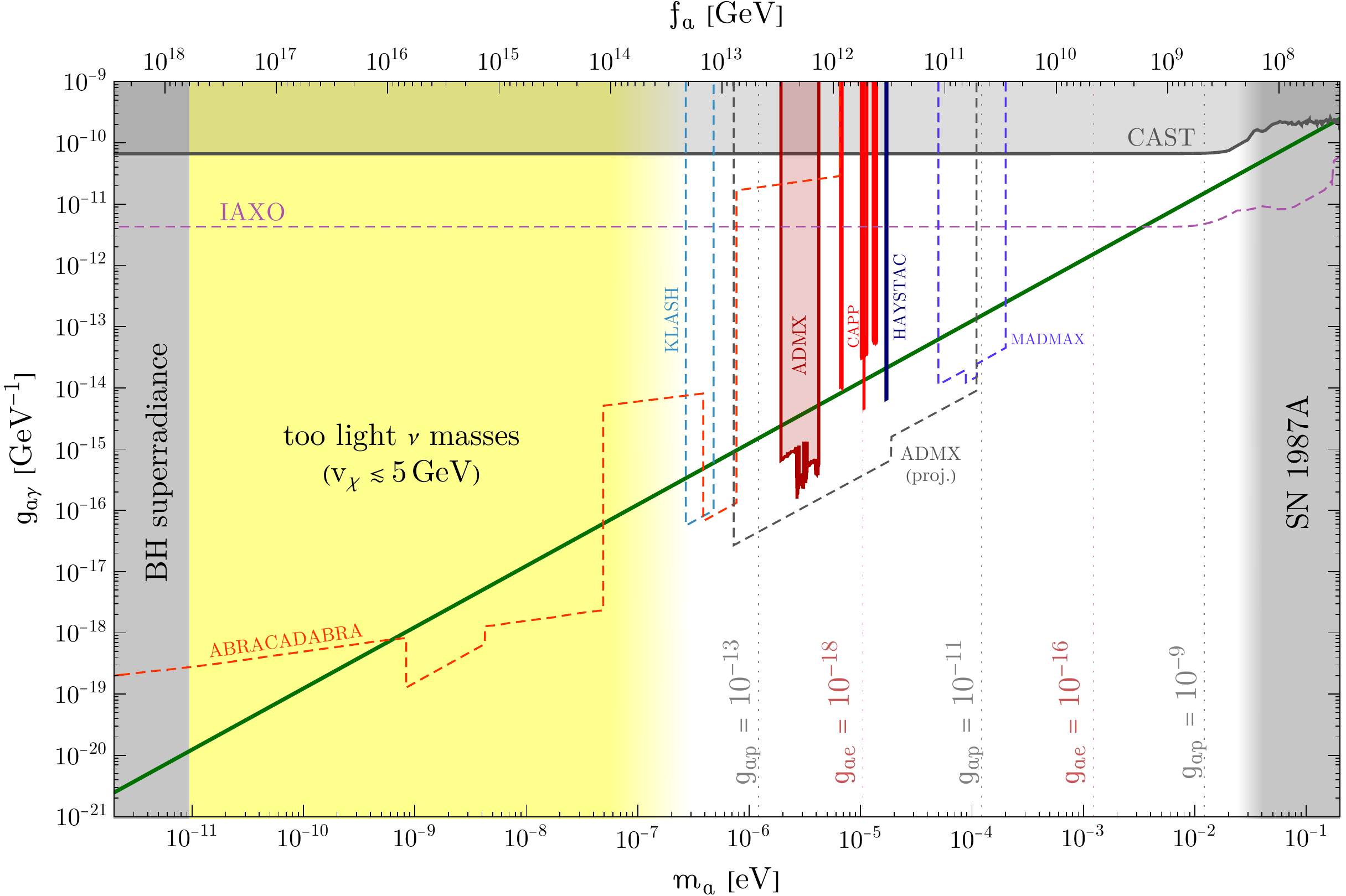}
\caption{
Allowed axion parameter space for the considered model, with $E/N=8$ (green line) and future detection prospects, in the plane of the axion mass $m_a$ vs.~the axion-photon coupling $g_{a\gamma}$. Vertical dotted lines show representative values of the axion-proton $g_{ap}$ (grey) and axion-electron $g_{ae}$ (red) couplings for this model.
Present data (CAST~\cite{CAST:2017uph}, ADMX~\cite{ADMX:2009iij,ADMX:2018gho,ADMX:2019uok,ADMX:2021nhd}, CAPP~\cite{Lee:2020cfj,Jeong:2020cwz,CAPP:2020utb}, HAYSTAC~\cite{HAYSTAC:2020kwv})
and projected sensitivities (IAXO~\cite{Shilon:2012te}, ABRACADABRA~\cite{Abracadabra}, KLASH~\cite{Alesini:2019nzq}, ADMX~\cite{Stern:2016bbw} and MADMAX~\cite{Beurthey:2020yuq}) are represented by solid and dashed lines, respectively (see also~\cite{DiLuzio:2020wdo,AxionLimits}).
}
\label{fig:axion}
\end{figure}

The electromagnetic anomaly coefficient $E$ is also independent of the charge $\alpha$. We find it is given by $E=2n_F$, where $n_F$ denotes the number of generations of vector-like seesaw partners $\DF$. We consider the minimal case (recall \cref{sec:fcontent}), for which $n_F = 2$ and $E =4$. 
This $E/N = 8$ value is safe from current axion experimental search bounds for large regions of the parameter space, as shown in \cref{fig:axion}. The allowed values of $f_a$ are constrained from below ($f_a \gtrsim 10^8$ GeV) due to the SN 1987A bound on the axion-nucleon couplings~\cite{Carenza:2019pxu}, which for our model implies the bound $g_{ap} \lesssim 3 \times 10^{-9}$ on the axion-proton coupling.
On the other hand, one expects an upper bound on $f_a$ from the relation in~\cref{eq:dss}. In particular, requiring perturbative Yukawa couplings of at most $\mathcal{O}(1)$ and taking a triplet VEV $v_\chi$ of at most $\mathcal{O}(5\text{ GeV})$, one sees that neutrino masses become suppressed beyond what is phenomenologically viable, i.e.~$m_\nu < \sqrt{\Delta m^2_\text{atm}} \simeq 0.05$ eV, unless $f_a \lesssim 10^{13-14}$ GeV.
Therefore, we take $f_a\in [10^8, 10^{13}]$ GeV as our viable range of interest.
As shown in~\cref{fig:axion}, large portions of the viable parameter space are expected to be probed by upcoming axion experiments.

\subsection{A heavier W mass?}
\label{sec:W}

Due to the engagement of $\DC$ in $SU(2)_L$ gauge interactions, the triplet extension of the Higgs sector can be constrained by electroweak precision measurements. In particular, a non-zero VEV $v_\chi$ in our model modifies the $\rho$ parameter, which at tree-level is calculated as 
\begin{equation}
\label{eq:rhotm}
	\rho \equiv \frac{M_W^2}{M_Z^2 \cos^2 \theta_W} = 1+4\frac{v^2_\chi}{v^2}\;,
\end{equation}
given the tree-level expressions for the squared masses of weak gauge bosons,
\begin{equation}
M_W^2 = \frac{g_2^2}{4} \left(v^2 + 4 v_\chi^2\right) \,, \qquad
M_Z^2 = \frac{g_2^2 \, v^2}{4 \cos^2 \theta_W} \,.
\end{equation}
Unlike in the SM case, custodial symmetry is not recovered in the limit $g' \to 0$.
We keep our discussion at the tree level in order to arrive at a plausible and illustrative benchmark for $v_\chi$. At this level, the scalar triplet VEV does not affect $M_Z$.

Significant attention has recently been given to models with hyperchargeless triplet scalars in light of the new $W$ mass measurement given by the CDF collaboration~\cite{CDF:2022hxs}
(see e.g.~\cite{Strumia:2022qkt,Asadi:2022xiy,DiLuzio:2022xns,PerezFileviez:2022gmy,Borah:2022obi,Popov:2022ldh,Batra:2022org,Cheng:2022aau,Addazi:2022fbj,Wang:2022dte,Evans:2022dgq,Lazarides:2022spe,Senjanovic:2022zwy,Ma:2022emu,Rizzo:2022jti}).%
\footnote{See also e.g.~\cite{Cheng:2022jyi,Kanemura:2022ahw,Heeck:2022fvl,Chakrabarty:2022voz,Bahl:2022gqg} for models with a $Y=1$ Higgs triplet addressing the CDF result.}
Taking the CDF II result as a hint for new physics, we re-express the $\rho$ parameter as
\begin{equation}
\label{eq:rhotmCDF}
	\rho = \frac{(M^{\mathrm{CDF}}_W)^2}{M^2_Z \cos^2\theta_W} \simeq \left(\frac{M^{\mathrm{CDF}}_W}{M^{\mathrm{SM}}_W}\right)^2 \rho_{\mathrm{SM}}\;,
\end{equation}
where $M_W^\mathrm{CDF} =80433.5\pm 9.4~\mathrm{MeV}$ is the CDF measurement, while the SM value is $ M_W^\mathrm{SM} =80357\pm 6~\mathrm{MeV}$~\cite{Zyla:2020zbs}.
Using central values, we obtain $\rho \simeq 1.0019$. Note that the change in $\rho$ has the correct positive sign in our model, at the tree level. \cref{eq:rhotm} then gives $v_\chi\simeq 5.36$ GeV. In the next section, we analyse the scalar potential of the model, taking $v_\chi = 5.4$ GeV as a benchmark value.

\section{Vacuum stability}
\label{sec:stability}
%
The real and complex scalars introduced in the previous section will have non-trivial effects on the vacuum structure. In this section, we analyse the vacuum structure in detail, taking the potential in \cref{eq:potential-sim} as our starting point, with an emphasis on electroweak vacuum stability.

\subsection{Mass spectrum}
\label{sec:mass_spectrum}
Assuming all the VEVs $(v,v_\chi, v_\sigma)$ are non-zero, we find the stationarity conditions as
\begin{subequations}
\label{eq:stationarity}
\begin{align}
\mu_H^2 &=\frac{1}{4} \left(2 \lambda v^2+ \lambda_a v_\chi^2 +\lambda_b v_\sigma^2 -2 \kappa v_\chi \right) \;,\\
\mu_\chi^2 &= \frac{1}{4 v_\chi} \left( \lambda_a v^2 v_\chi+ \lambda_c v_\sigma^2 v_\chi+\lambda_\chi  v_\chi^3 - \kappa v^2 \right) \;,\\
\mu_\sigma^2 &= \frac{1}{4} \left(  \lambda_b v^2+2 \lambda_\sigma  v_\sigma^2+\lambda_c v_\chi^2 \right) \;,
\end{align}
\end{subequations}
under the parameterisations $H=(\phi^+,(\phi^0+v+ i G_H)/\sqrt{2})^T$, $\sigma=(\phi_\sigma + v_\sigma + i G_\sigma)/\sqrt{2}$. With these conditions, we find the mass matrix for the neutral scalars in the basis ($\phi^0,\chi^0,\phi_\sigma$) to be
\begin{align}
M^2_{\text{CP-even}}=
\begin{pmatrix}
 \lambda v^2 & \frac{1}{2}\lambda_a v v_\chi-\frac{1}{2} \kappa v & \frac{1}{2} \lambda_b v v_\sigma \\[4mm]
 \frac{1}{2}\lambda_a v v_\chi-\frac{1}{2} \kappa v & \frac{\kappa  v^2}{4v_\chi}+\frac{\lambda_\chi v_\chi^2}{2} & \frac{1}{2} \lambda_c v_\sigma v_\chi \\[4mm]
 \frac{1}{2} \lambda_b v v_\sigma & \frac{1}{2} \lambda_c v_\sigma v_\chi & \lambda_\sigma v_\sigma^2 
 \end{pmatrix}\,,
\label{eq:CP-even-mass-matrix}
\end{align}
The VEVs $(v,v_\chi,v_\sigma)$ correspond to a minimum of the potential in \cref{eq:potential-sim} when this $M^2_{\text{CP-even}}$ matrix is positive definite. The CP-odd mass matrix vanishes, corresponding to two Goldstone bosons. One of them becomes the longitudinal component of $Z$ boson and the other one is the axion, which becomes massive when the chiral axion potential is considered, as in the KSVZ model.

The three CP-even mass eigenstates have masses $m_{H_i}$ ($i=1,2,3$). These obey $m_{H_1} \sim m_{H_2} \ll m_{H_3}$, where the last one is much larger than the first two due to the large $v_\sigma$ and the corresponding eigenstate effectively decouples.
Indeed, from the considerations of the previous section, we have $v_\chi/v_\sigma \lesssim 10^{-8}$, which indicates the hierarchy of the VEVs, $v_\chi < v \ll v_\sigma$. As a result, we further decompose the $3\times 3$ mass matrix into four blocks
\begin{align}
M^2_{\text{CP-even}}=
\begin{pmatrix}
M_{h}^2 & M_{h\sigma}^2 \\[2mm]
(M_{h\sigma}^2)^T &  M_\sigma^2 
\end{pmatrix}\,,
\label{eq:decompose-mass-matrix}
\end{align}
with
\begin{align}
    M_{h}^2 \equiv
\begin{pmatrix}
 \lambda v^2 & \frac{1}{2}\lambda_a v v_\chi-\frac{1}{2} \kappa v\\[4mm]
 \frac{1}{2}\lambda_a v v_\chi-\frac{1}{2} \kappa v & \frac{\kappa  v^2}{4v_\chi}+\frac{\lambda_\chi v_\chi^2}{2}  
 \end{pmatrix}\,, \quad
   M_{h\sigma}^2 \equiv 
\begin{pmatrix}
 \frac{1}{2} \lambda_b v v_\sigma \\[4mm]
 \frac{1}{2} \lambda_c v_\sigma v_\chi
\end{pmatrix}\,, \quad
M_\sigma^2 \equiv \lambda_\sigma v_\sigma^2 \,.
\end{align}
The masses of the neutral scalars receive contributions from couplings to $\sigma$. In the limit of vanishing $M_h^2$, the couplings to $\sigma$ contribute to the $2\times 2$ $M_h^2$ block in the diagonalisation and can be calculated in a seesaw-like approximation as
\begin{align}
    (M_h^{\sigma})^2 \equiv-\frac{1}{M_\sigma^2} M_{h\sigma}^2 (M_{h\sigma}^2)^T = -\frac{1}{4\lambda_\sigma}
\begin{pmatrix}
 \lambda_b^2 v^2 & \displaystyle\lambda_b \lambda_c v v_\chi \\[2mm]
 \lambda_b \lambda_c v v_\chi & \displaystyle \lambda_c^2 v_\chi^2
\end{pmatrix}\,.
\end{align}
A rough estimate tells us that this contribution is of the same order as that of $M_h^2$. Consequently, the leading-order mass matrix for the two light scalars $(\phi^0,\chi^0)$ reads
\begin{align}
M_{h\chi}^2 \equiv M_h^2+(M_h^{\sigma})^2  = 
\begin{pmatrix}
 \lambda  v^2-\frac{\lambda_b^2 v^2}{4 \lambda_\sigma } & -\frac{\kappa  v}{2}+\frac{\lambda_a v v_\chi}{2}-\frac{\lambda_b \lambda_c v v_\chi}{4 \lambda_\sigma } \\[4mm]
 -\frac{\kappa  v}{2}+\frac{\lambda_a v v_\chi}{2}-\frac{\lambda_b \lambda_c v v_\chi}{4 \lambda_\sigma } & \frac{\kappa  v^2}{4 v_\chi}-\frac{\lambda_c^2 v_\chi^2}{4 \lambda_\sigma }+\frac{\lambda_\chi  v_\chi^2}{2} 
\end{pmatrix}\,,
\label{eq:m2x2}
\end{align}
which leads to an estimate for the neutral scalar mixing angle $\alpha$ of
\begin{align}
\tan 2\alpha = \frac{2 v v_\chi (2 \kappa  \lambda_\sigma -2 \lambda_a \lambda_\sigma  v_\chi +\lambda_b \lambda_c v_\chi)}{v_\chi^3 \left(\lambda_c^2-2 \lambda_\sigma  \lambda_\chi \right)-v^2 \left[\kappa  \lambda_\sigma +v_\chi \left(\lambda_b^2-4 \lambda  \lambda_\sigma \right)\right]} \;,\label{eq:mixing}
\end{align}
such that $O^T M_{h\chi}^2O=\mathrm{Diag}\{(m_{H_1}^\mathrm{LO})^2,(m_{H_2}^\mathrm{LO})^2\}$ with $O$ being a $2\times 2$ rotation matrix parameterised by the angle $\alpha$.
This approximation will be useful in understanding the parameter correlations discussed in \cref{sec:numerical}. While one can solve for $m_{H_1}^\mathrm{LO}$ and $m_{H_2}^\mathrm{LO}$ starting from \cref{eq:m2x2}, the expressions are lengthy and we do not show them here. 
Although the block-diagonalisation procedure discussed so far is convenient to understand the leading-order contributions, in our numeric study we take into account the full $3\times 3$ matrix as given in \cref{eq:CP-even-mass-matrix}. 

The mass matrix for the charged scalars in the basis $(\phi^\pm, \chi^\pm)$ is
\begin{equation} 
M^2_{\mathrm{charged}} = 
\begin{pmatrix}
 \kappa v_\chi & \displaystyle\frac{\kappa  v}{2} \\[4mm]
 \frac{\kappa v}{2} & \displaystyle\frac{\kappa  v^2}{4 v_\chi}
\end{pmatrix}\,.
\label{eq:charged-mass-matrix}
\end{equation} 
One of the two charged-scalar masses is zero, corresponding to the charged Goldstone boson that becomes the longitudinal component of $W^\pm$. The only non-zero squared mass is
\begin{align}
    m_{H^\pm}^2 &= \frac{\kappa  \left( v^2+4 v_\chi^2\right)}{4 v_\chi}\,,
\label{eq:mhpm}
\end{align}
which grows with $\kappa$.
The mixing is given by $\tan 2\beta =4 v v_\chi/(v^2-4 v_\chi^2)$.
Inputting $v_\chi$ constrained by the CDF result, we find a value of $\beta\simeq 0.044$ for the mixing angle.

\subsection{Constraints}
The model parameter space is subject to many constraints. To start with, the potential should be bounded from below in any direction of large field values. This condition can be quantified by requiring the copositivity of the quartic coupling matrix~\cite{Kannike:2012pe}. In our model, the copositivity conditions read
\begin{subequations}
\begin{align}
    &\lambda \ge 0\,,\quad \lambda_\chi \ge 0\,,\quad \lambda_\sigma \ge 0\,, \\[1mm]
    & \lambda_a+\sqrt{2 \lambda  \lambda_\chi }\ge 0\,,\quad
    \lambda_b + 2 \sqrt{\lambda  \lambda_\sigma }\ge 0\,,\quad
    \lambda_c +\sqrt{2 \lambda_\sigma  \lambda_\chi }\ge 0\,,   \\[1mm] 
    & \lambda_a \sqrt{2\lambda_\sigma }+\lambda_b \sqrt{\lambda_\chi }+\lambda_c \sqrt{2\lambda }  +2 \sqrt{\lambda \lambda_\chi \lambda_\sigma}\nonumber \\ 
    & +\sqrt{2\left(\lambda_a + \sqrt{2\lambda  \lambda_\chi }\right) \left(\lambda_b +2 \sqrt{\lambda  \lambda_\sigma }\right)  \left( \lambda_c  +  \sqrt{2\lambda_\chi \lambda_\sigma }\right)}\ge 0\,.
\end{align}
\end{subequations}
The perturbativity bound requires instead that all the quartic couplings remain perturbative at any scale, i.e.~$|\lambda_i| < 4\pi$ (with $i$ being a pseudo-index running over all the quartic couplings).
There are also constraints from requiring the unitarity of the $S$-matrix~\cite{Khan:2016sxm,Chabab:2018ert}:
\begin{align}
     |\lambda|,~ |\lambda_\sigma| < 8\pi, \quad |\lambda_a|,~ |\lambda_b|,~ |\lambda_c|,~ |\lambda_\chi| < 16\pi \;.
\end{align}
Unitarity gives three additional constraints, namely upper bounds on the quantities in \cref{eq:eigvm20} of \cref{apd:unitarity}, where more details on the unitarity bounds can be found.

Additionally, we are interested in identifying the regions of parameter space where the desired vacuum configuration $(v,v_\chi,v_\sigma)$ with all VEVs non-zero is a global minimum. We therefore need to exclude deeper minima from alternative configurations $(v',v_\chi',v_\sigma')$ with one or more vanishing VEVs.
Confining our attention to such charge-conserving VEVs, a direct minimum depth comparison results in a difference $\Delta V = V' - V$ with
\begin{equation}
\begin{aligned}
\Delta V \,&=\, \frac{1}{16} \Big[
2 \lambda \left(v^4-v'^4\right)
+ \lambda_\chi \left(v_\chi^4-v_\chi'^4\right)
+2 \lambda_\sigma \left( v_\sigma^4-v_\sigma'^4\right)
+2 \lambda_c \left( v_\chi^2 v_\sigma^2  - v_\chi'^2 v_\sigma'^2 \right) \\
&\qquad\quad+2 v^2 \left(\lambda_a v_\chi^2 + \lambda_b v_\sigma^2 - \kappa  v_\chi\right)
-2 v'^2 \left(\lambda_a v_\chi'^2 + \lambda_b v_\sigma'^2 - \kappa  v_\chi'\right)
   \Big]\,,
\label{eq:vc}
\end{aligned}
\end{equation}
which we require to be non-negative for all seven patterns $(v',v_\chi',v_\sigma') = (0,0,0)$, $(v',0,0)$, $(0,v_\chi',0)$, $(0,0,v_\sigma')$, $(v',v_\chi',0)$, $(v',0,v_\sigma')$, $(0,v_\chi',v_\sigma')$, in case these lead to a positive definite mass-squared matrix.
Note that the primed VEVs in this equation are constrained to satisfy stationarity conditions of their own, but with the same $\mu_H^2$, $\mu_\chi^2$ and $\mu_\sigma^2$ as given in \cref{eq:stationarity} in terms of unprimed VEVs.
We also check that the minima candidates are not locally destabilised by turning on charge-breaking VEVs (see~\cite{Ferreira:2019hfk} for an in-depth analysis in the $Y=1$ triplet case).

Constraints from the experimental side arise mainly from two sources: electroweak precision measurements and collider experiments. For the former, we consider the constraint on the triplet VEV $v_\chi$ from the $\rho$ parameter, taking into account the latest measurement of the $W$ mass (see \cref{sec:W}). The bounds on oblique parameters, i.e.~on the Peskin-Takeuchi parameters $S$, $T$, and $U$~\cite{Peskin:1991sw}, also impose stringent constraints on models of new physics above the electroweak scale. The contributions beyond the tree level from the hyperchargeless triplet to a modified version of these parameters, adapted to this context, are~\cite{Forshaw:2001xq,Forshaw:2003kh}
\begin{subequations}
\label{eq:oblique}
\begin{align}
    S&\simeq 0 \;,\\
    T&= \frac{1}{8\pi} \frac{1}{\sin^2 \theta_W \cos^2 \theta_W} \left[ \frac{m_{H_2}^2+m_{H^\pm}^2}{m_Z^2} - \frac{2 m_{H^\pm}^2 m_{H_2}^2}{m_Z^2 (m_{H_2}^2-m_{H^\pm}^2)} \mathrm{log} \left( \frac{m_{H_2}^2}{m_{H^\pm}^2}\right)\right] \nonumber\\
    &\simeq \frac{1}{6\pi} \frac{1}{\sin^2 \theta_W \cos^2 \theta_W} \frac{(\Delta m)^2}{m_Z^2} \;,\\
    U&\simeq -\frac{1}{3\pi} \left[ m_{H_2}^4 \mathrm{log} \left( \frac{m_{H_2}^2}{m_{H^\pm}^2}\right) \frac{3 m_{H^\pm}^2- m_{H_2}^2}{(m_{H_2}^2-m_{H^\pm}^2)^3} +\frac{5( m_{H^\pm}^4+ m_{H_2}^4)-22 m_{H_2}^2m_{H^\pm}^2}{6(m_{H_2}^2-m_{H^\pm}^2)^2}\right]\nonumber\\
    &\simeq \frac{\Delta m}{3\pi m_{H^\pm}} \;,
\end{align}
\end{subequations}
where $m_Z$ is the $Z$ boson mass, $\theta_W$ is the Weinberg angle, and $\Delta m \equiv m_{H_2}- m_{H^\pm}$. The last approximations hold for $|\Delta m| \ll m_{H^\pm}$.
To be consistent with the updated fits of Refs.~\cite{Strumia:2022qkt,Lu:2022bgw}, we require that $m_{H^\pm} \sim m_{H_2}\gg m_{H_1}$, where $m_{H_1}$ is assumed to be the SM Higgs. 
To be more specific, given that the benchmark $v_\chi = 5.4$ GeV already produces a large tree-level contribution $T_\text{tree} = \beta^2/\alpha_\text{e.m.} \simeq 0.25$, the fitted CDF value of $T = 0.27 \pm 0.06$ \cite{Lu:2022bgw} (under the assumption of negligible $U$) requires the additional loop-level contributions of \cref{eq:oblique} to be small, leading to the bound $|\Delta m| < 50$ GeV.

As for collider constraints, an important channel is that of Higgs decay into two photons, corresponding to a signal strength of $\mu_{\gamma\gamma}=1.14^{+0.19}_{-0.18}$~\cite{ATLAS:2016neq}. In our model, the novel contribution to $\mu_{\gamma\gamma}$ is dominated by $\lambda_a/m_{H^\pm}^2$ and can be made negligible if $m_{H^\pm}>300$ GeV~\cite{Khan:2016sxm}.  
Additionally, LEP provides the most stringent bound on the mass of a neutral scalar which is produced in association with the $Z$ boson, $m_h>114$ GeV~\cite{OPAL:2000ghn,DELPHI:2004bco}. However, it is possible to evade this bound in a hyperchargeless triplet model since the coupling of the new neutral scalar to the $Z$ may be suppressed~\cite{Chabab:2018ert}. Therefore, in case there is a scalar lighter than the SM Higgs, we impose the constraint
\begin{align}
|\sin \alpha| < 0.05\;, \label{eq:light-H-bound}
\end{align}
which implies that $|\cos \alpha| \simeq 1$ and the LEP bound is not violated.
If the lightest scalar has a mass below half the SM Higgs mass, it can contribute to the Higgs invisible decay rate and is subject to further constraints. Meanwhile, $\chi^0$ also couples to the $W$ boson and has the potential to be produced via such an interaction. A full analysis of the parameter space taking into consideration all Higgs search limits is beyond the scope of the current work. In a simplified analysis, we focus on \cref{eq:light-H-bound} as the main constraint on a light scalar spectrum.
Finally, the vector-like fermion triplets $\DF$ acquire masses proportional to the PQ symmetry breaking scale and are thus safe from otherwise stringent low-energy limits (see e.g.~\cite{CMS:2013czn}).

\subsection{Numerical results and discussion}
\label{sec:numerical}

Following the analysis of the mass spectrum and the above discussion on constraints, we are ready to search for the viable parameter space at the electroweak scale. We express the potential parameters $\mu_H^2,~\mu_\chi^2,~\mu_\sigma^2$ in terms of the non-zero VEVs $(v,v_\chi,v_\sigma)$ and the quartic and trilinear couplings using the stationarity conditions in~\cref{eq:stationarity}. The constraints to the quartic couplings can be directly applied. Meanwhile, there are constraints expressed in terms of the masses, which are also functions of the VEVs and the quartic and trilinear couplings, according to \cref{eq:CP-even-mass-matrix,eq:charged-mass-matrix}. 
There are two possible mass spectra with some differences in constraints ($m_H$ is the SM Higgs mass):
\begin{itemize}
    \item ``Heavy spectrum'', with $m_{H_1} = m_H < m_{H_2}$, referring to the case where the new scalar has mass $m_{H_2}$ and is heavier than the SM Higgs. In this case, the oblique parameter constraints of \cref{eq:oblique} require $m_{H^\pm} \sim m_{H_2}\gg m_{H}$, as mentioned before.
    \item ``Light spectrum'', with $m_{H_1}<m_{H_2}=m_H$, referring to the case where the new scalar has mass $m_{H_1}$ and is lighter than the SM Higgs. Oblique parameter constraints, which assume the scale of new physics to be large with respect to the electroweak scale, do not apply. Instead, we consider the bound of \cref{eq:light-H-bound} to suppress the coupling to the $Z$ boson such that the LEP bound is not violated.
\end{itemize}
Numerically, we fix the VEVs to be 
$$ v=246~\mathrm{GeV},\quad v_\sigma=10^{12}~\mathrm{GeV}, \quad v_\chi=5.4~\mathrm{GeV} \;,$$
and randomly scan the trilinear and the quartic couplings in the ranges
$$\kappa \in [10,100]~\mathrm{GeV}, \quad \lambda, \lambda_\chi, \lambda_\sigma, |\lambda_a|,|\lambda_b|, |\lambda_c| \in [0,4\pi] \;.$$
In practice, for quartic couplings we take a flat logarithmic prior with a lower limit of $10^{-6}$.
We also require that the  mass of the SM Higgs-like scalar lies in the $3\sigma$ range of $125.25\pm 0.51$~GeV~\cite{Zyla:2020zbs}, and scan for both possibilities, $m_{H_1}=m_H<m_{H_2}$ and $m_{H_1}<m_{H_2}=m_H$. 

\begin{table}[t]
\centering
\begin{tabular}{ccc}
\toprule
      & $\,\,\,$ Light spectrum $(m_{H_2}=m_H)\,\,\,$ & $\,\,\,$ Heavy spectrum $(m_{H_1}=m_H)\,\,\,$ \\
     \midrule
     $\lambda$    &    $[0.0011,0.26]$            & $[0.26,1.22]$  \\[2mm] 
     $\lambda_\chi$ &  $[1.01\times 10^{-6},0.10]$            & $[1.01\times 10^{-6},10.89]$  \\[2mm]
     $\lambda_\sigma$& $[1.05\times 10^{-6},8.99\times 10^{-4}]$          & $[1.08\times 10^{-6},11.53]$  \\[2mm]
     $\lambda_a$ &     $[0.03,3.15]$               & $[-1.27,12.14]$  \\[2mm]
     $\lambda_b$ &     $[-0.0044,-0.0001]$               & $[1.11\times 10^{-6},5.40]$ \\[2mm]
     $\lambda_c$ &     $[0.10,3.68]$               & $[1.34\times 10^{-6},12.17]$ \\[2mm] 
     $\kappa$ [GeV]&   $[32.06,99.93]$          & $[32.07,99.90]$ \\[2mm]
     \midrule
     $m_{H_1}(m_{H_2})$ [GeV] & $[0.25,123.15]$      &$[258.34, 529.29]$ \\[2mm]
     $m_{H_3}$ [GeV]         &  $[1.00\times 10^9,3.00\times 10^{10}]$      &$[1.04\times 10^9,3.40\times 10^{12}]$\\[2mm]
     $m_{H^\pm}$ [GeV]       &  $[300.01, 529.63]$     &$[300.04, 529.55]$ \\[2mm]
\bottomrule
\end{tabular}
\caption{The viable parameter space at the electroweak scale.}
\label{tab:numeric-ew}
\end{table}

The parameter ranges for points satisfying both the theoretical and experimental constraints are shown in \cref{tab:numeric-ew}.
For these points, we will further check if they allow the desired vacuum $(v, v_\chi, v_\sigma)$ to be the global one. This is done by numerically checking whether the potential defined by the scanned parameters admits other types of vacua. If so, we compare the depth of the latter with that of the desired vacuum to guarantee they are not deeper, see \cref{eq:vc}.
For the heavy spectrum, we find that the desired vacuum can be the global one, while for the light spectrum, it may only be a local one. In particular, for a light spectrum, the vacuum of the type $(0,v_\chi^\prime,0)$ is always deeper than the desired one. The difference $\Delta V$ is 
\begin{align}
V_{(0,v_\chi^\prime,0)} - V_{(v,v_\chi,v_\sigma)} &= \frac{1}{16} \left[ 2 \lambda  v^4+2 v^2 \left(\lambda_b v_\sigma^2+\lambda_a v_\chi^2-\kappa  v_\chi\right)+2 \lambda_\sigma  v_\sigma^4+2 \lambda_c v_\sigma^2 v_\chi^2+\lambda_\chi  \left(v_\chi^4-v_\chi^{\prime 4}\right)\right] \nonumber\\
& \simeq \frac{1}{16} \left(2 \lambda_\sigma  v_\sigma^4 -\lambda_\chi v_\chi^{\prime 4} \right)\,.
\end{align}
Numerically, we find the points passing all the other constraints lead to $v_\chi^\prime \sim \mathcal{O}(10^{13})$ GeV, and thus to a negative value of the difference, given positive $\lambda_\sigma$ and $\lambda_\chi$.
Although not being stable, it is still possible that the desired vacuum is meta-stable in the sense that the tunnelling time to other, deeper vacua is longer than the age of the Universe. We do not investigate this possibility here.

We require the couplings to remain perturbative and that the desired vacuum stays stable up to the PQ breaking scale, where other new physics is expected to come in. The evolution of the couplings is governed by the one-loop renormalisation group equations (RGEs), which are calculated using SARAH~\cite{Staub:2013tta,Staub:2015kfa}, see \cref{sec:apd-RGE}.
As a first approximation, we analyse the RGE-improved tree-level potential (see also e.g.~\cite{Zhang:2015uuo,Ghorbani:2021rgs}). 
The parameter space shown in \cref{tab:numeric-ew} gets further constrained by perturbativity, copositivity and unitarity at the PQ scale. Roughly speaking, large values of the quartic couplings are ruled out. 

\begin{figure}[t]
    \centering
    \includegraphics[width=0.9\textwidth]{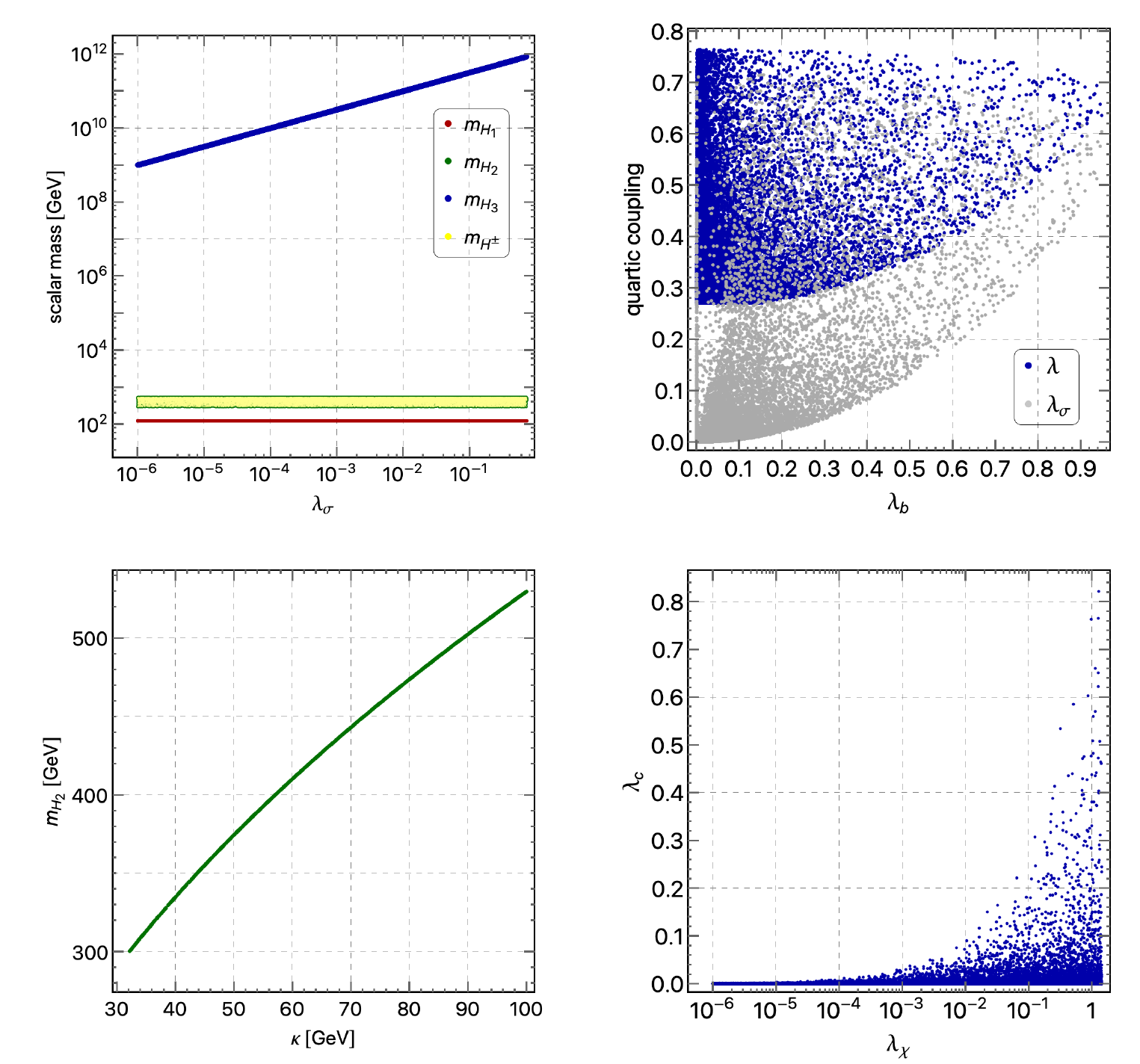}
    \caption{Two-dimensional projections of the viable parameter space for the heavy spectrum, $m_{H_1}=m_H<m_{H_2}$, satisfying all the constraints while having the desired vacuum as a global minimum.
    }
    \label{fig:correlation-h}
\end{figure}

\begin{figure}[t]
    \centering
    \includegraphics[width=0.9\textwidth]{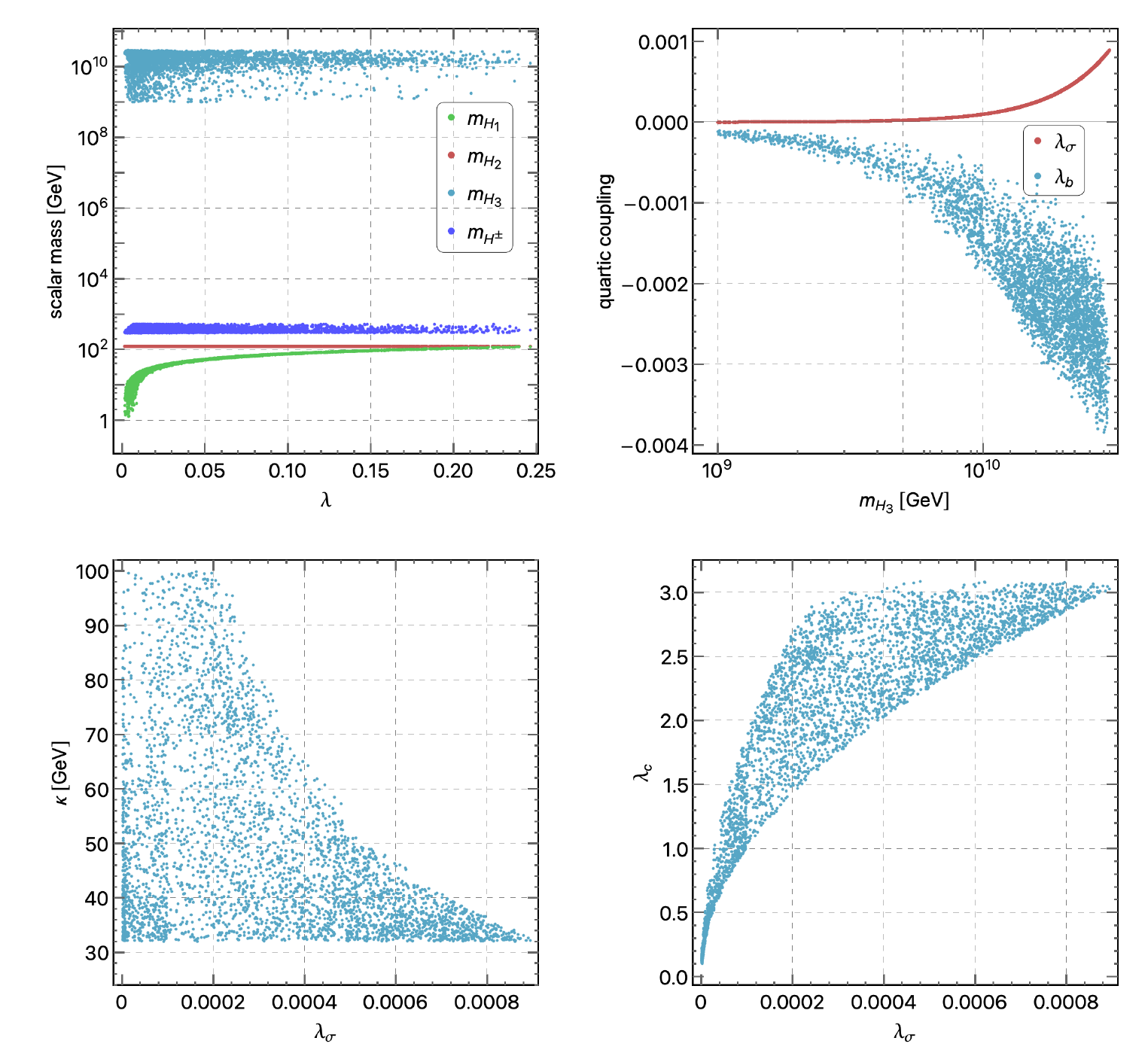}
    \caption{
    Two-dimensional projections of the viable parameter space for the light spectrum, $m_{H_1}<m_{H_2}=m_H$, satisfying all the constraints, but with the desired vacuum as a local minimum (see text).}
    \label{fig:correlation-l}
\end{figure}

The final viable parameter space of our model is presented in the form of two-parameter projections in \Cref{fig:correlation-h,fig:correlation-l}.
For both spectra, we find regions of viable parameter space at the PQ scale, meaning that the desired vacuum can be stable at least up to this scale. For the heavy spectrum (\cref{fig:correlation-h}) several comments are in order:
\begin{itemize}
    \item 
    The top-left plot in \cref{fig:correlation-h} shows the mass spectrum with varying $\lambda_\sigma$.
    As one may expect from the discussion of \cref{sec:mass_spectrum}, the approximate relation $m_{H_3}^2 \simeq \lambda_\sigma v_\sigma^2$ holds.        
    Other scalar masses do not seem to be sensitive to $\lambda_\sigma$, even after taking into account all the discussed constraints, especially those on the mass spectrum (matching the SM Higgs mass and satisfying the upper limit on the charged-neutral mass splitting $\Delta m$).
    The upper bound of $m_{H_3}$ is set by the upper bound of $\lambda_\sigma$. There is no lower bound on $m_{H_3}$, which can be made smaller at the cost of tuning $\lambda_\sigma$ to very small values.
    \item 
    The top-right plot shows the correlations between the Higgs quartic $\lambda$ and the $\sigma$-related quartic couplings $\lambda_b$ and $\lambda_\sigma$. These are not affected by the requirement of having a global minimum.
     The bottom-left plot shows instead the dependence of the scalar mass $m_{H_2}$ on $\kappa$. Since the charged-scalar mass $m_{H^\pm}$ depends solely on $\kappa$ (see \cref{eq:mhpm}) and $|\Delta m|$ is bounded, $m_{H_2}$ is expected to grow with $\sqrt{\kappa}$. 
     We find this dependence becomes rather sharp, i.e.~the numerically allowed values of mass splitting become quite small ($|\Delta m| \lesssim 0.1$ GeV), after excluding points not leading to a global minimum.
The bottom-right plot shows the relation between $\lambda_\chi$ and $\lambda_c$. The $\lambda_\chi$-dependent upper bound on $\lambda_c$ arises only after applying the global minimum filter.
    \item 
    Comparing the plotted parameter ranges with those in \cref{tab:numeric-ew}, we see that all the quartic coupling ranges shrink. Indeed, running up to the PQ scale and imposing the relevant constraints at that scale excludes the large-quartic portion of the parameter space, as previously mentioned. As we have seen in the previous point, asking for a global minimum also imposes non-trivial restrictions. This requirement further excludes the region $\lambda_a < 0.5$ for points with $\Delta m < 0$, and the upper bound on $\lambda_c$ becomes more stringent overall, going from $\lambda_c \lesssim 3$ to $\lambda_c \lesssim 0.8$.
\end{itemize}

For the light spectrum (\cref{fig:correlation-l}), we also plot the final viable parameter space in terms of its projections in planes of two parameters, and several comments are in order:
\begin{itemize}
    \item
    The top-left plot in \cref{fig:correlation-l} shows the mass spectrum with varying $\lambda$. Only $m_{H_1}$ grows with $\lambda$ while the other masses are mostly insensitive to it. In contrast to the case of the heavy spectrum, the heaviest neutral scalar $m_{H_3}$ now spans a much narrower range, roughly from $10^9$ GeV to $10^{10}$ GeV, which results from a much more stringent upper bound on $\lambda_\sigma$ ~(cf.~\cref{{tab:numeric-ew}} and the other subplots in the figure).
    \item 
    The top- and bottom-right plots involve parameters directly related to $\sigma$ and can be read together to understand the upper limit on $m_{H_3}$. The mass $m_{H_3}$ grows with $\lambda_\sigma$ and thus with $|\lambda_b|$ and $\lambda_c$, due to the their correlations. However, $\lambda_c$ grows with $\lambda_\sigma$ rather fast, easily leading to exclusion when the RGE running is accounted for, bounding $\lambda_\sigma$ and consequently $m_{H_3}$ from above. The dependence of $m_{H_3}$ on $\lambda_\sigma$ is otherwise similar to that of the heavy mass spectrum and has no lower bound if we allow $\lambda_\sigma$, $|\lambda_b|$ and $\lambda_c$ to be vanishingly small.
    \item 
    The bottom-left plot shows the relation between $\kappa$ and $\lambda_\sigma$. The lower bound on $\kappa$ is set by the lower bound on the charged Higgs mass. Besides the chosen cut at $100$ GeV, we see that there is an upper bound on $\kappa$, which becomes more stringent for larger $\lambda_\sigma$. This rough bound can be understood from the formula of \cref{eq:mixing} for the light scalar mixing and the requirement that said mixing is small, see \cref{eq:light-H-bound}. In particular, $\tan 2\alpha$ depends on $\kappa$ via the product $\kappa\lambda_\sigma$, which
    determines the observed exclusion.
\end{itemize}

\vfill
\clearpage
\section{Conclusions}
\label{sec:summary}
%
In this work, we investigate the connection between tree-level Dirac neutrino masses and axion physics in a scenario where the PQ scale $f_a$ is the only heavy scale to play a role in neutrino mass generation. To realise such a connection, we focus on the diagram of \Cref{fig:feynman} as the main contribution to neutrino masses and build the model based on it. The minimal construction leads us to a KSVZ-type model, in which the SM scalar sector is extended by a real triplet $\Delta_\chi$ and by the PQ field $\sigma$. Scalars other than $\sigma$ are not charged under PQ. We find the PQ symmetry by itself is enough to explain the Dirac nature of neutrino masses in such a setup, i.e.~the PQ symmetry enforces lepton number conservation perturbatively.

The scale $f_a$ suppresses Dirac neutrino masses and is consequently bounded from above, $f_a \lesssim 10^{13}$ GeV. The QCD axion in the model addresses the strong CP problem, while being a potential dark matter candidate. Future prospects for its detection have been discussed (see \cref{fig:axion}). In turn, the real scalar triplet contributes to the $W$ boson mass via its VEV and may be responsible for the recent hint of beyond-the-SM physics by the CDF collaboration. Finally, we look into the scalar sector of the model. We identify the regions in parameter space compatible with the desired VEV structure, taking into account electroweak precision constraints and the requirements of copositivity and perturbativity up to the PQ scale.
Besides the SM-like Higgs, there is another light neutral scalar that can be either heavier or lighter than the former. The two possible spectra are dubbed ``heavy'' and ``light'', respectively. We find that for the heavy spectrum the desired EW vacuum can be the global one up to the PQ scale, while it is only found to be a local one in the case of the light spectrum.

This work can be extended in many ways. On the one hand, a full survey of the light-spectrum parameter space, as well as the heavy-spectrum one, may lead to interesting collider phenomenology. On the other hand, there are rich Yukawa structures to be explored, which, working e.g.~with flavour symmetry, have the potential to address the neutrino mixing pattern and enhance the predictive power. Last but not least, it would be interesting to examine whether neutrinogenesis is viable in this context.

\vspace{0.5cm}
\section*{Acknowledgements}

X.Y.Z.~would like to thank Prof.~Shun Zhou for helpful discussions and comments.
The work of J.T.P.~was supported by
Fundação para a Ciência e a Tecnologia (FCT, Portugal) through the projects
PTDC/FIS-PAR/29436/2017, 
CERN/FIS-PAR/0004/2019, CERN/FIS-PAR/\linebreak[0]0008/2019, and
CFTP-FCT Unit 777 (namely UIDB/00777/2020 and UIDP/00777/2020),
which are partially funded through POCTI (FEDER), COMPETE, QREN and EU.
The work of Y.R.~was supported by the Doctoral Program of Tian Chi Foundation of Xinjiang Uyghur Autonomous Region of China under grant No.~TCBS202128 and by the Natural Science Foundation of Xinjiang Uyghur Autonomous Region of China under grant No.~2022D01C52.
The work of X.Y.Z.~was supported in part by the National Natural Science Foundation of China under grant No.~11835013 and by the Key Research Program of the Chinese Academy of Sciences under grant No.~XDPB15.

\vfill
\clearpage

\appendix
\section{Unitarity bounds on the quartic couplings}\label{apd:unitarity}
The unitarity of the scattering matrix for $2\to 2$ process puts constraints on the model parameters. At high energies, according to the Goldstone boson equivalence theorem, scattering amplitudes of the longitudinal gauge boson can be well approximated by those of the corresponding Goldstone boson. Dominant contributions to the scattering amplitudes come from the quartic couplings of the scalars. In the following, we compute all possible $2\to 2$ scattering matrices, classified by the total charges of the initial/final state particles.

Considering the total electric charge of the initial states is zero, the S-matrix can be written as a direct sum of the following two matrices:
\begin{equation}
\mathcal{M}^{(0)}_1 = \left(
\begin{array}{cccccccccccc}
	\lambda  & 0 & 0 & 0 & 0 & 0 & 0 & 0 & 0 & 0 & 0 & 0 \\
	0 & \lambda_\sigma   & 0 & 0 & 0 & 0 & 0 & 0 & 0 & 0 & 0 & 0 \\
	0 & 0 & \frac{\lambda_a}{2} & 0 & 0 & 0 & 0 & 0 & 0 & 0 & 0 & 0 \\
	0 & 0 & 0 & \frac{\lambda_a}{2} & 0 & 0 & 0 & 0 & 0 & 0 & 0 & 0 \\
	0 & 0 & 0 & 0 & \frac{\lambda_a}{2} & 0 & 0 & 0 & 0 & 0 & 0 & 0 \\
	0 & 0 & 0 & 0 & 0 & \frac{\lambda_a}{2} & 0 & 0 & 0 & 0 & 0 & 0 \\
	0 & 0 & 0 & 0 & 0 & 0 & \frac{\lambda_b}{2} & 0 & 0 & 0 & 0 & 0 \\
	0 & 0 & 0 & 0 & 0 & 0 & 0 & \frac{\lambda_b}{2} & 0 & 0 & 0 & 0 \\
	0 & 0 & 0 & 0 & 0 & 0 & 0 & 0 & \frac{\lambda_b}{2} & 0 & 0 & 0 \\
	0 & 0 & 0 & 0 & 0 & 0 & 0 & 0 & 0 & \frac{\lambda_b}{2} & 0 & 0 \\
	0 & 0 & 0 & 0 & 0 & 0 & 0 & 0 & 0 & 0 & \frac{\lambda_c}{2} & 0 \\
	0 & 0 & 0 & 0 & 0 & 0 & 0 & 0 & 0 & 0 & 0 & \frac{\lambda_c}{2} \\
\end{array}
\right)\,,
\end{equation}
for the initial state basis $\left(\phi^0 G_H,\phi_\sigma G_\sigma, \chi^0 G_H, \phi^0 \chi^0, \phi^+ \chi^-, \chi^+ \phi^-, \phi^0 G_\sigma, \phi^0 \phi_\sigma, \phi_\sigma G_H, G_H G_\sigma, \chi^0 G_\sigma, \phi_\sigma \chi^0\right)$,
and
\begin{equation}
\mathcal{M}^{(0)}_2 = \left(
\begin{array}{ccccccc}
	\frac{3 \lambda }{2} & \frac{\lambda_b }{4} & \frac{\lambda }{2} & \frac{\lambda_b }{4} & \frac{\lambda }{\sqrt{2}} & \frac{\lambda_a }{2 \sqrt{2}} & \frac{\lambda_a }{4} \\
	\frac{\lambda_b }{4} & \frac{3 \lambda_\sigma  }{2} & \frac{\lambda_b }{4} & \frac{\lambda_\sigma  }{2} & \frac{\lambda_b }{2 \sqrt{2}} & \frac{\lambda_c }{2 \sqrt{2}} & \frac{\lambda_c }{4} \\
	\frac{\lambda }{2} & \frac{\lambda_b }{4} & \frac{3 \lambda }{2} & \frac{\lambda_b }{4} & \frac{\lambda }{\sqrt{2}} & \frac{\lambda_a }{2 \sqrt{2}} & \frac{\lambda_a }{4} \\
	\frac{\lambda_b }{4} & \frac{\lambda_\sigma  }{2} & \frac{\lambda_b }{4} & \frac{3 \lambda_\sigma  }{2} & \frac{\lambda_b }{2 \sqrt{2}} & \frac{\lambda_c }{2 \sqrt{2}} & \frac{\lambda_c }{4} \\
	\frac{\lambda }{\sqrt{2}} & \frac{\lambda_b }{2 \sqrt{2}} & \frac{\lambda }{\sqrt{2}} & \frac{\lambda_b }{2 \sqrt{2}} & 2 \lambda  & \frac{\lambda_a }{2} & \frac{\lambda_a }{2 \sqrt{2}} \\
	\frac{\lambda_a }{2 \sqrt{2}} & \frac{\lambda_c }{2 \sqrt{2}} & \frac{\lambda_a }{2 \sqrt{2}} & \frac{\lambda_c }{2 \sqrt{2}} & \frac{\lambda_a }{2} & \lambda_\chi   & \frac{\lambda_\chi  }{2 \sqrt{2}} \\
	\frac{\lambda_a }{4} & \frac{\lambda_c }{4} & \frac{\lambda_a }{4} & \frac{\lambda_c }{4} & \frac{\lambda_a }{2 \sqrt{2}} & \frac{\lambda_\chi  }{2 \sqrt{2}} & \frac{3 \lambda_\chi  }{4} \\
\end{array}
\right)\,,
\end{equation}
for the initial states $\left(\phi^0 \phi^0/\sqrt{2}, \phi_\sigma \phi_\sigma/\sqrt{2}, G_H G_H/\sqrt{2}, G_\sigma G_\sigma/\sqrt{2}, \phi^+ \phi^-, \chi^+ \chi^-, \chi^0 \chi^0/\sqrt{2}\right)$.
Here, the factor $1/\sqrt{2}$ takes care of the statistics for identical particles. The eigenvalues of the matrix $\mathcal{M}^{(0)}_2$ are $\lambda$ (with multiplicity 2), $\lambda_\sigma$, $\lambda_\chi/2$, and
\begin{equation}\label{eq:eigvm20}
  \frac{1}{12} A +\frac{1}{2}\sqrt{\frac{B}{3}}\cos \left[\frac{1}{3}\cos^{-1} \left( \frac{3C}{2}\sqrt{\frac{3}{B^3}} \right) +\frac{2k\pi}{3}\right]\,, \quad \text{with }k=0,1,2\,,
\end{equation}
where
\begin{subequations}
\begin{align}
    A & = 12\lambda + 8\lambda_\sigma+5\lambda_\chi\,,\\
    B & = \frac{A^2}{3} -96 \lambda  \lambda_\sigma -60 \lambda  \lambda_\chi +12 \lambda_a^2+8 \lambda_b^2+6 \lambda_c^2-40 \lambda_\sigma  \lambda_\chi\,,\\
    C & = \frac{A}{3}\left( B-\frac{A^2}{9}   \right)\,.
\end{align}
\end{subequations}
For the charge $+ 1$ initial states $\left(\phi^+ \phi^0, \phi^+ G_H, \chi^+ \chi^0, \phi^+ \chi^0, \chi^+ \phi^0, \chi^+ G_H, \phi^+ G_\sigma, \phi^+ \phi_\sigma, \chi^+ \phi_\sigma, \chi^+ G_\sigma\right)$, we obtain the following $10\times 10$
diagonal S-matrix
\begin{equation}
  \mathcal{M}^{(+1)} =  \left(
\begin{array}{cccccccccc}
	\lambda  & 0 & 0 & 0 & 0 & 0 & 0 & 0 & 0 & 0 \\
	0 & \lambda  & 0 & 0 & 0 & 0 & 0 & 0 & 0 & 0 \\
	0 & 0 & \frac{\lambda_\chi }{2} & 0 & 0 & 0 & 0 & 0 & 0 & 0 \\
	0 & 0 & 0 & \frac{\lambda_a}{2} & 0 & 0 & 0 & 0 & 0 & 0 \\
	0 & 0 & 0 & 0 & \frac{\lambda_a}{2} & 0 & 0 & 0 & 0 & 0 \\
	0 & 0 & 0 & 0 & 0 & \frac{\lambda_a}{2} & 0 & 0 & 0 & 0 \\
	0 & 0 & 0 & 0 & 0 & 0 & \frac{\lambda_b}{2} & 0 & 0 & 0 \\
	0 & 0 & 0 & 0 & 0 & 0 & 0 & \frac{\lambda_b}{2} & 0 & 0 \\
	0 & 0 & 0 & 0 & 0 & 0 & 0 & 0 & \frac{\lambda_c}{2} & 0 \\
	0 & 0 & 0 & 0 & 0 & 0 & 0 & 0 & 0 & \frac{\lambda_c}{2} \\
\end{array}
\right)\,.
\end{equation}
Finally, the S-matrix of the charge $+ 2$ initial states $\left(\phi^+ \phi^+/\sqrt{2}, \chi^+ \chi^+/\sqrt{2}, \phi^+ \chi^+\right)$ is
\begin{equation}
  \mathcal{M}^{(+2)} = \left(
\begin{array}{ccc}
	\lambda  & 0 & 0 \\
	0 & \frac{\lambda_\chi }{2} & 0 \\
	0 & 0 & \frac{ \lambda_a}{2} \\
\end{array}
\right)\,.
\end{equation}
 Unitarity constraints on the S-matrices force the absolute values of the eigenvalues of the matrices to be less than $8\pi$. This implies the following upper bounds on the quartic couplings
 \begin{equation}
     |\lambda|,~ |\lambda_\sigma| < 8\pi\,, 
     \quad |\lambda_a|,~ |\lambda_b|,~ |\lambda_c|,~ |\lambda_\chi| < 16\pi\,,
\end{equation}
and, on top of these, extra constraints are imposed by bounding the eigenvalues of the matrix $\mathcal{M}^{(0)}_2$,  given in \cref{eq:eigvm20}.

\section{One-loop RGEs}\label{sec:apd-RGE}
In this work, we calculate the RGEs up to the one-loop level using the {\it Mathematica} package SARAH~\cite{Staub:2013tta,Staub:2015kfa}. The beta function of the coupling $X$ is defined as
\begin{align}
    \beta_X=\mu\frac{\partial X}{\partial \mu} = \frac{1}{16\pi^2} \beta_X^{(1)}\ \;.
\end{align}
The beta functions for the gauge couplings read
\begin{subequations}
\begin{align} 
\beta_{g_1}^{(1)} & =  
\frac{41}{10} g_{1}^{3} \;,\\ 
\beta_{g_2}^{(1)} & =  
\frac{5}{2} g_{2}^{3} \;, \\ 
\beta_{g_3}^{(1)} & =  
-\frac{19}{3} g_{3}^{3} \;.
\end{align}
\end{subequations}
Here, $g_1 \equiv \sqrt{5/3}\, g'$. Note that the (high-energy) beta function for $g_2$ is modified not just by the new scalar triplet but also by the new triplet fermions $\DF$. The beta function of $g_3$ is instead modified with respect to the SM one due to the presence of the vector-like quark $Q$.
The beta functions for the Yukawa couplings are
\begin{subequations}
\begin{align} \allowdisplaybreaks
\beta_{Y_{F}}^{(1)}  =&~  
\frac{1}{8} \Big(2 {Y_{F}  Y_{F}^{\dagger}  Y_{F}}  + 2 {Y_F  Y_{L}^{\dagger}  Y_L}  + 6 Y_{F} \Big(-16 g_{2}^{2}  + 4 |Y_Q|^2  + \mbox{Tr}\Big({Y_{F}  Y_{F}^{\dagger}}\Big)\Big) + {Y_{R}^{T}  Y_{R}^*  Y_{F}}\Big)\;,\\ 
\beta_{Y_L}^{(1)}  =&~  
\frac{1}{8} \Big(5 \Big(4 {Y_e  Y_{e}^{\dagger}  Y_L}  + {Y_L  Y_{L}^{\dagger}  Y_L}\Big) + {Y_L  Y_{F}^{\dagger}  Y_F}\Big)\nonumber \\ 
 &+Y_L \Big(3 \mbox{Tr}\Big({Y_d  Y_{d}^{\dagger}}\Big)  + 3 \mbox{Tr}\Big({Y_u  Y_{u}^{\dagger}}\Big)  -\frac{33}{4} g_{2}^{2}  + \frac{3}{4} \mbox{Tr}\Big({Y_L  Y_{L}^{\dagger}}\Big)  -\frac{9}{20} g_{1}^{2}  + \mbox{Tr}\Big({Y_e  Y_{e}^{\dagger}}\Big)\Big)\;,\\ 
\beta_{Y_u}^{(1)}  =&~  
\frac{3}{2} \Big({Y_u  Y_{u}^{\dagger}  Y_u}  - {Y_d  Y_{d}^{\dagger}  Y_u}\Big)+Y_u \Big(3 \mbox{Tr}\Big({Y_d  Y_{d}^{\dagger}}\Big)  + 3 \mbox{Tr}\Big({Y_u  Y_{u}^{\dagger}}\Big) \nonumber \\ 
 & -8 g_{3}^{2}  -\frac{17}{20} g_{1}^{2}  + \frac{3}{4} \mbox{Tr}\Big({Y_L  Y_{L}^{\dagger}}\Big)  -\frac{9}{4} g_{2}^{2}  + \mbox{Tr}\Big({Y_e  Y_{e}^{\dagger}}\Big)\Big)\;,\\ 
\beta_{Y_Q}^{(1)}  =&~  
4 Y_{Q}^{2} Y_Q^*  -8 g_{3}^{2} Y_Q  + \frac{3}{4} Y_Q \mbox{Tr}\Big({Y_{F}  Y_{F}^{\dagger}}\Big) \;,\\ 
\beta_{Y_{R}}^{(1)}  =&~  
\frac{1}{8} {Y_{R}  Y_{F}^*  Y_{F}^{T}}  + Y_{R} \Big(-6 g_{2}^{2}  + \frac{1}{2} \mbox{Tr}\Big({Y_{R}  Y_{R}^{\dagger}}\Big) \Big) + {Y_{R}  Y_{R}^{\dagger}  Y_{R}}\;,\\ 
\beta_{Y_d}^{(1)}  =&~  
\frac{1}{4} \Big(6 \Big({Y_d  Y_{d}^{\dagger}  Y_d}- {Y_u  Y_{u}^{\dagger}  Y_d} \Big)+Y_d \Big(12 \mbox{Tr}\Big({Y_d  Y_{d}^{\dagger}}\Big)  + 12 \mbox{Tr}\Big({Y_u  Y_{u}^{\dagger}}\Big)-32 g_{3}^{2}  \nonumber \\ 
 & + 3 \mbox{Tr}\Big({Y_L  Y_{L}^{\dagger}}\Big)  + 4 \mbox{Tr}\Big({Y_e  Y_{e}^{\dagger}}\Big)  -9 g_{2}^{2}  - g_{1}^{2} \Big)\Big)\;,\\ 
\beta_{Y_e}^{(1)}  =&~  
\frac{1}{8} \Big(3 \Big(4 {Y_e  Y_{e}^{\dagger}  Y_e}  + 5 {Y_L  Y_{L}^{\dagger}  Y_e} \Big)+Y_e \Big(24 \mbox{Tr}\Big({Y_d  Y_{d}^{\dagger}}\Big) \nonumber \\ 
 & + 6 \Big(-3 \Big(g_{1}^{2} + g_{2}^{2}\Big) + 4 \mbox{Tr}\Big({Y_u  Y_{u}^{\dagger}}\Big)  + \mbox{Tr}\Big({Y_L  Y_{L}^{\dagger}}\Big)\Big) + 8 \mbox{Tr}\Big({Y_e  Y_{e}^{\dagger}}\Big) \Big)\Big)\;.
\end{align}
\end{subequations}
The beta functions for the quartic scalar couplings are
\begin{subequations}
\begin{align} \allowdisplaybreaks
\beta_{\lambda}^{(1)}  =&~  
\frac{27}{100} g_{1}^{4} +\frac{9}{10} g_{1}^{2} g_{2}^{2} +\frac{9}{4} g_{2}^{4} -\frac{9}{5} g_{1}^{2} \lambda -9 g_{2}^{2} \lambda +12 \lambda^{2} +\frac{3}{4} {\lambda}_{a}^{2} +\frac{1}{2} {\lambda}_{b}^{2} +12 \lambda \mbox{Tr}\Big({Y_d  Y_{d}^{\dagger}}\Big) \nonumber \\ 
 &+4 \lambda \mbox{Tr}\Big({Y_e  Y_{e}^{\dagger}}\Big) +3 \lambda \mbox{Tr}\Big({Y_L  Y_{L}^{\dagger}}\Big) +12 \lambda \mbox{Tr}\Big({Y_u  Y_{u}^{\dagger}}\Big) -12 \mbox{Tr}\Big({Y_d  Y_{d}^{\dagger}  Y_d  Y_{d}^{\dagger}}\Big) -4 \mbox{Tr}\Big({Y_e  Y_{e}^{\dagger}  Y_e  Y_{e}^{\dagger}}\Big) \nonumber \\ 
 &-4 \mbox{Tr}\Big({Y_e^{\dagger}  Y_L  Y_L^{\dagger}  Y_e}\Big) -\frac{5}{4} \mbox{Tr}\Big({Y_L  Y_{L}^{\dagger}  Y_L  Y_{L}^{\dagger}}\Big) -12 \mbox{Tr}\Big({Y_u  Y_{u}^{\dagger}  Y_u  Y_{u}^{\dagger}}\Big) \;,\\ 
\beta_{\lambda_{\sigma}}^{(1)}  =&~  
10 \lambda_{\sigma}^{2}  + 12 \lambda_{\sigma} |Y_Q|^2  -12 |Y_Q|^4  + 3 \lambda_{\sigma} \mbox{Tr}\Big({Y_{F}  Y_{F}^{\dagger}}\Big)  + \frac{3}{4} {\lambda}_{c}^{2}  -\frac{3}{4} \mbox{Tr}\Big({Y_{F}  Y_{F}^{\dagger}  Y_{F}  Y_{F}^{\dagger}}\Big)  + {\lambda}_{b}^{2}\\ 
\beta_{{\lambda}_{\chi}}^{(1)}  =&~ -24 g_{2}^{2} {\lambda}_{\chi}  + 2 {\lambda}_{a}^{2}  + 2 {\lambda}_{\chi} \mbox{Tr}\Big({Y_{R}  Y_{R}^{\dagger}}\Big)  -2 \mbox{Tr}\Big({Y_{R}  Y_{R}^{\dagger}  Y_{R}  Y_{R}^{\dagger}}\Big)  + \frac{19}{2} {\lambda}_{\chi}^{2}  + {\lambda}_{c}^{2}\;,\\ 
\beta_{{\lambda}_{b}}^{(1)}  =&~  
-\frac{9}{10} g_{1}^{2} {\lambda}_{b} -\frac{9}{2} g_{2}^{2} {\lambda}_{b} +6 \lambda {\lambda}_{b} +2 {\lambda}_{b}^{2} +\frac{3}{2} {\lambda}_{a} {\lambda}_{c} +4 {\lambda}_{b} \lambda_{\sigma} +6 {\lambda}_{b} |Y_Q|^2 +6 {\lambda}_{b} \mbox{Tr}\Big({Y_d  Y_{d}^{\dagger}}\Big) \nonumber \\ 
 &~+2 {\lambda}_{b} \mbox{Tr}\Big({Y_e  Y_{e}^{\dagger}}\Big) +\frac{3}{2} {\lambda}_{b} \mbox{Tr}\Big({Y_{F}  Y_{F}^{\dagger}}\Big) +\frac{3}{2} {\lambda}_{b} \mbox{Tr}\Big({Y_L  Y_{L}^{\dagger}}\Big) +6 {\lambda}_{b} \mbox{Tr}\Big({Y_u  Y_{u}^{\dagger}}\Big) -\frac{3}{2} \mbox{Tr}\Big({Y_{F}^{\dagger}Y_F  Y_L^{\dagger}  Y_L}\Big) \;,\\ 
\beta_{{\lambda}_{a}}^{(1)}  =&~ 
-\frac{9}{10} g_{1}^{2} {\lambda}_{a} -\frac{33}{2} g_{2}^{2} {\lambda}_{a} +6 \lambda {\lambda}_{a} +2 {\lambda}_{a}^{2} +{\lambda}_{b} {\lambda}_{c} +\frac{5}{2} {\lambda}_{a} {\lambda}_{\chi} +6 {\lambda}_{a} \mbox{Tr}\Big({Y_d  Y_{d}^{\dagger}}\Big) +2 {\lambda}_{a} \mbox{Tr}\Big({Y_e  Y_{e}^{\dagger}}\Big) \nonumber \\ 
&~+\frac{3}{2} {\lambda}_{a} \mbox{Tr}\Big({Y_L  Y_{L}^{\dagger}}\Big) +{\lambda}_{a} \mbox{Tr}\Big({Y_{R}  Y_{R}^{\dagger}}\Big) +6 {\lambda}_{a} \mbox{Tr}\Big({Y_u  Y_{u}^{\dagger}}\Big) \;,\\ 
\beta_{{\lambda}_{c}}^{(1)}  =&~  
2 {\lambda}_{a} {\lambda}_{b} -12 g_{2}^{2} {\lambda}_{c} +2 {\lambda}_{c}^{2} +4 {\lambda}_{c} \lambda_{\sigma} +\frac{5}{2} {\lambda}_{c} {\lambda}_{\chi} +6 {\lambda}_{c} |Y_Q|^2 +\frac{3}{2} {\lambda}_{c} \mbox{Tr}\Big({Y_{F}  Y_{F}^{\dagger}}\Big) +{\lambda}_{c} \mbox{Tr}\Big({Y_{R}  Y_{R}^{\dagger}}\Big) \nonumber \\ 
&~- \mbox{Tr}\Big({Y_{R}^{\dagger}  Y_{R}  Y_{F}^*  Y_{F}^{T}}\Big) \;.
\end{align}
\end{subequations}
Finally, the beta function for the trilinear scalar coupling is
\begin{align}  \allowdisplaybreaks
\beta_{\kappa}^{(1)}  =&~  
\frac{1}{10} \kappa \Big(-9 g_{1}^{2} -105 g_{2}^{2} +20 \lambda +20 {\lambda}_{a} +60 \mbox{Tr}\Big({Y_d  Y_{d}^{\dagger}}\Big) +20 \mbox{Tr}\Big({Y_e  Y_{e}^{\dagger}}\Big) +15 \mbox{Tr}\Big({Y_L  Y_{L}^{\dagger}}\Big)  \nonumber \\ 
 &~+5 \mbox{Tr}\Big({Y_{R}  Y_{R}^{\dagger}}\Big)+60 \mbox{Tr}\Big({Y_u  Y_{u}^{\dagger}}\Big) \Big)\;.
\end{align}

\bibliographystyle{JHEPwithnote}
\bibliography{bibliography}

\end{document}